\documentstyle[prl,aps]{revtex}
\draft

\begin{document}
\input{epsf}
\twocolumn[\hsize\textwidth\columnwidth\hsize\csname@twocolumnfalse\endcsname
\title{Digital Communication Using Chaotic Pulse Generators}
\author{Nikolai F. Rulkov$^{1}$, Mikhail M. Sushchik$^{1}$,
Lev S. Tsimring$^{1}$, Alexander R. Volkovskii$^{1}$,Henry D.I.
Abarbanel$^{1,2}$, Lawrence Larson$^{3}$, and  Kung Yao$^{4}$}
\address{
 $^1$ Institute for Nonlinear Science, UCSD, La Jolla, CA
92093-0402, USA \\
 $^2$ Marine Physical Laboratory, Scripps Institute of Oceanography and Department of
Physics, University of California, San Diego, La Jolla, CA
92093-0402, USA\\
 $^3$ Electrical and Computer Engineering Department, UCSD, La Jolla, CA
92093-0407, USA\\
 $^4$ Electrical Engineering Department, UCLA, Los Angeles,
  CA 90095-1594, USA}

\date{\today}
\maketitle

\begin{abstract}
Utilization of chaotic signals for covert communications remains a
very promising practical application. Multiple studies indicated
that the major shortcoming of recently proposed chaos-based
communication schemes is their susceptibility to noise and
distortions in communication channels. In this paper we discuss a
new approach to communication with chaotic signals, which
demonstrates good performance in the presence of channel
distortions. This communication scheme is based upon chaotic
signals in the form of pulse trains where intervals between the
pulses are determined by chaotic dynamics of a pulse generator. The
pulse train with chaotic interpulse intervals is used as a carrier.
Binary information is modulated onto this carrier by the pulse
position modulation method, such that each pulse is either left
unchanged or delayed by a certain time, depending on whether ``0''
or ``1'' is transmitted. By synchronizing the receiver to the
chaotic pulse train we can anticipate the timing of pulses
corresponding to ``0'' and ``1'' and thus can decode the
transmitted information. Based on the results of theoretical and
experimental studies we shall discuss the basic design principles
for the chaotic pulse generator, its synchronization, and the
performance of the chaotic pulse communication scheme in the
presence of channel noise and filtering.

\end{abstract}
\narrowtext
\vskip2pc]

\section{Introduction}

Noise-like signals generated by deterministic systems with chaotic
dynamics have a high potential for many applications including
communication. Very rich, complex and flexible structure of such
chaotic signals is the result of local instability of
post-transient motions in a generator of chaos. This is achieved as
a result of specific features of nonlinear vector field in the
phase space of the generator and not by increasing the design
complexity. Even a simple nonlinear circuit with very few
off-the-shelf electronic components is capable of generating a very
complex set of chaotic signals. The simplicity of chaos generators
and the rich structure of chaotic signals are the most attractive
features of deterministic chaos that have caused a significant
interest in possible utilization of chaos in communication.

Chaotic signals exhibit a broad continues spectrum and have been
studied in connection with spread-spectrum
applications~\cite{Mazzini98}. Due to their irregular nature, they
can be used to efficiently encode the information in a number of
ways. Thanks to the deterministic origin of the chaotic signals two
coupled chaotic systems can be synchronized to produce identical
chaotic oscillations\cite{Pecora90}. This provides the key to
recovery of information that is modulated onto a chaotic
carrier\cite{Wu93}.  A number of chaos-based covert communication
schemes have been suggested\cite{Hasler98inls}, but many of these
are very sensitive to distortions, filtering, and
noise\cite{Chen98inls,Rulkov99}. The negative effect of filtering
is primarily due to the extreme sensitivity of nonlinear systems to
phase distortions.  This limits the use of filtering for noise
reduction in chaos-based communications.  One way to avoid this
difficulty is to use chaotically timed pulse sequences rather than
continuous chaotic waveforms\cite{Rulkov93}.  Each pulse has
identical shape, but the time delay between them varies
chaotically. Since the information about the state of the chaotic
system is contained {\em entirely} in the timing between pulses,
the distortions that affect the pulse shape will not significantly
influence the ability of the chaotic pulse generators to
synchronize and thus be utilized in communications.  This proposed
system is similar to other ultra-wide bandwidth impulse radios
\cite{Win98} that offers a very promising communication platform,
especially in severe multi-path environments or where they are
required to co-exist with a large number of other radio systems.
Chaotically varying the spacing between narrow pulses enhances the
spectral characteristics of the system by removing any periodicity
from the transmitted signal. Because of the absence of
characteristic frequencies, chaotically positioned pulses are
difficult to observe and detect for the unauthorized user.  Thus
one expects that transmission based on chaotic pulse sequences can
be designed to have a very low probability of intercept.

In this paper we discuss the design of a self-synchronizing
chaos-based impulse communication system, and present the results
of the performance analysis in the demonstration setup operating
through a model channel with noise, filtering, and attenuation. We
consider the case where the encoding of the information signal is
based upon the alteration of time position of pulses in the chaotic
train. This encoding method is called Chaotic Pulse Position
Modulation~\cite{Sushchik99}

\section{Chaotic Pulse Position Modulation}
In this section we describe the method of Chaotic Pulse Position
Modulation (CPPM) and basic elements of its hardware
implementation. Consider a chaotic pulse generator which produces
chaotic pulse signal
\begin{equation}\label{signal}
U(t)=\sum_{j=0}^{\infty}w(t-t_j),
\end{equation}
where $w(t-t_j)$ represents the  waveform of a pulse generated at
time $t_j=t_0+\sum_{n=0}^{j}T_n$, and $T_n$ is the time interval
between the $n$-th and $(n-1)$-th pulses. We assume that the
sequence of the time intervals, $T_i$, represents iterations of a
chaotic process. For simplicity we will consider the case where
chaos is produced by a one-dimensional map $T_n= F(T_{n-1})$, where
$F(~)$ is a nonlinear function. Some studies on such chaotic pulse
generators can be found in~\cite{Rulkov93,Bernhard92}.

Using the Chaotic Pulse Position Modulation method the information
is encoded within the chaotic pulse signal by using additional
delays in the generated interpulse intervals, $T_n$. As a result,
the generated pulse sequence is given by a new map of the form
\begin{equation}\label{map_enc}
T_n= F(T_{n-1})+d+mS_n,
\end{equation}
where $S_n$ is the information-bearing signal. Here we will
consider only the case of binary information, and therefore, $S_n$
equals to zero or one. The parameter $m$ characterizes the
amplitude of modulation. The parameter $d$ is a constant time delay
which is needed for practical implementation of our modulation and
demodulation method. The role of this parameter will be specified
later. In the design of chaotic pulse generator the nonlinear
function $F(~)$, and parameters $d$ and $m$ are selected to
guarantee chaotic behavior of the map.

The modulated chaotic pulse signal
$U(t)=\sum_{j=0}^{\infty}w(t-t_0-\sum_{n=0}^{j}T_n)$, where $T_n$
is generated by Eq.(\ref{map_enc}) is the transmitted signal. The
duration of each pulse $w(t)$ in the pulse train is assumed to be
much shorter than the minimal value of the interpulse intervals,
$T_n$. To detect information at the receiver end, the decoder is
triggered by the received pulses, $U(t)$. The consecutive time
intervals $T_{n-1}$ and $T_{n}$ are measured and the information
signal is recovered from the chaotic iterations $T_n$ with the
formula
\begin{equation}\label{map_dec}
S_n= (T_n-F(T_{n-1})-d)/m,
\end{equation}
If the nonlinear function, $F(~)$, and parameters $d$ and $m$ in
the authorized receiver are the same as in the transmitter, then
the encoded information, $S_n$, can be easily recovered. When the
nonlinear functions are not matched with sufficient precision, a
large decoding error results. In other words, an unauthorized
receiver has no information on the spacing between the pulses in
the transmitted signal, it cannot determine whether a particular
received pulse was delayed, and thus whether $S_n$ was ``0'' or
``1''.

Since the chaotic map of the decoder in the authorized receiver is
matched to the map of the encoder in the corresponding transmitter,
the time of the next arriving pulse can be predicted. In this case
the input of the synchronized receiver can be blocked up to the
moment of time when the next pulse is expected. The time intervals
when the input to a particular receiver is blocked can be utilized
by other users, thus providing a multiplexing strategy. Such
selectivity helps to improve the performance of the system by
reducing the probability of false triggering of the decoder by
channel noise.

\begin{figure}[hbt]
\begin{center}
\leavevmode
\hbox{%
\epsfxsize=8.0cm
\epsffile{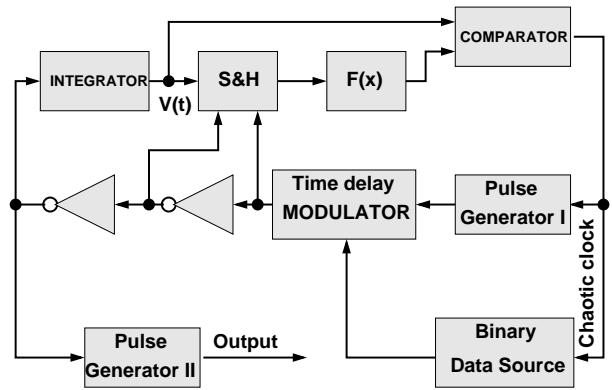}}
\end{center}
\caption{Block diagram of the chaotic pulse modulator.}
\label{modulator}
\end{figure}

\subsection{Transmitter}

The implementation of the chaotic pulse modulator used in our
experiments is illustrated in Fig.\ref{modulator}. The Integrator
produces a linearly increasing voltage, $V(t)=\beta^{-1}(t-t_n)$,
at its output. At the Comparator this voltage is compared with the
threshold voltage produced at the output of the nonlinear converter
$F(x)$. The threshold level $F(V_n)$ is formed by a nonlinear
conversion of voltage $V_n=V(t_n)$ which was acquired and saved
from the previous iteration using sample and hold (S\&H) circuits.
When voltage $V(t)$ reaches this threshold level, the comparator
triggers the Pulse Generator I. It happens at the moment of time
$t'_{n+1}=t_n+\beta F(V_n)$. The generated pulse (Chaotic Clock
Signal) causes the Data Generator to update the transmitted
information bit. Depending on what the information bit $S_{n+1}$ is
being transmitted, the Delay Modulator delays the pulse produced by
the Pulse Generator by the time $d+mS_{n+1}$. Therefore the delayed
pulse is generated at the moment of time $t_{n+1}=t_n+\beta F(V_n)
+d+mS_{n+1}$. Through the sample and hold circuit (S\&H) this pulse
first resets the threshold to the new iteration value of the
chaotic map $V(t_{n+1})\rightarrow F(V(t_{n+1}))$, and then resets
the integrator output to zero, $V(t)=0$. The dynamics of the
threshold is determined by the shape nonlinear function $F(~)$. The
spacing between the $n$-th and $(n+1)$-th pulses is proportional to
the threshold value $V_n$, which is generated according to the map
\begin{equation}\label{map_enc_exp}
T_{n+1}=\beta F(\beta^{-1}T_n)+d+mS_{n+1},
\end{equation}
where $T_n=t_{n-1}-t_n$, and $S_n$ is the binary information
signal. In the experimental setup the shape of the nonlinear
function was built to have the following form
\begin{equation}
F(x)\equiv \alpha f(x)=\cases{
\alpha x & \mbox{if $x < 5V$}, \cr
\alpha(10V-x) & \mbox{if $x \geq 5V$}.}
\label{nlf}
\end{equation}
The selection of the nonlinearity in the form of piece-wise linear
function helps to ensure the robust regimes of chaos generation for
rather broad ranges of parameters of the chaotic pulse position
modulator.

The position-modulated pulses, $w(t-t_j)$ are shaped in the Pulse
Generator II. These pulses form the output signal $U(t)=
\sum_{j=0}^{\infty}w(t-t_j)$, which is transmitted to the receiver.

\subsection{Receiver}

\begin{figure}[hbt]
\begin{center}
\leavevmode
\hbox{%
\epsfxsize=8.0cm
\epsffile{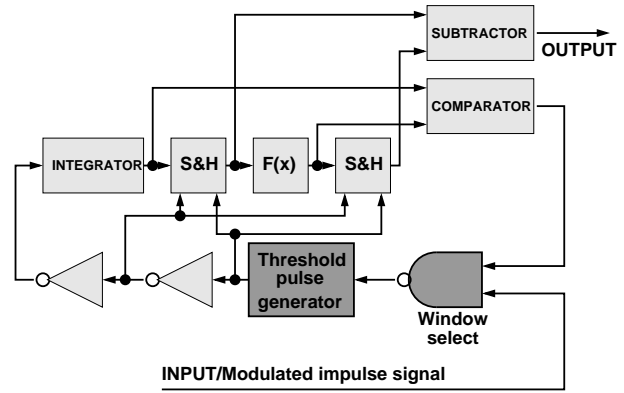}}
\end{center}
\caption{Block diagram of the chaotic pulse demodulator.}
\label{demodulator}
\end{figure}

When the demodulator is synchronized to the pulse position
modulator, then in order to decode a single bit of transmitted
information we must determine whether a pulse from the transmitter
was or was not delayed relative to its anticipated position. If an
ideal synchronization is established, but the signal is corrupted
by noise, the optimal detection scheme operates as follows.
Integrate the signal over the pulse duration inside the windows
where pulses corresponding to ``1'' and ``0'' are expected to
occur. The decision on whether ``1'' or ``0'' is received is made
based upon whether the integral over ``1''-window is larger or
smaller than that over ``0''-window. Such detection scheme in the
ideal case of perfect synchronization is the ideal Pulse Position
Modulation (PPM) scheme.  The performance of this scheme is known
to be 3dB worse than the BPSK system. Although in the case of
perfect synchronization this detection scheme is ideal, according
to our numerical simulations, its performance quickly degrades when
synchronization errors due to the channel noise are taken into
account. For this reason and for the sake of design simplicity we
use a different approach to detection.  The demodulator scheme is
illustrated in Fig.\ref{demodulator}.

In the receiver the Integrator, S\&H circuits and the nonlinear
function block generating the threshold values are reset or
triggered by the pulse received from the transmitter rather than by
the pulse from the internal feedback loop. To be more precise, they
are triggered when the input signal, $U(t)$, from the channel
exceeds certain input threshold. The time difference between the
anticipated location of the pulse without modulation,
$t'_{n+1}=t_n+\beta F(V_n)$,  and the actual arrival time $t_{n+1}$
translates into the difference between the threshold value,
$F(V_n)$ generated by the nonlinear function and the voltage,
$V(t_{n+1})$ at the Integrator at the moment when the input signal,
$U(t)$ exceeds the input threshold. For each received pulse the
difference $V(t_{n+1})-F(V_n)$ is computed and is used for deciding
whether or not the pulse was delayed. If this difference is less
than the reference value $\beta (d+m/2)$, the detected data bit
$S_{n+1}$ is ``0'', otherwise it is ``1''.

Another important detail of the receiver is the Window Selection
block. Once the receiver correctly observes two consecutive pulses,
it can predict the earliest moment of time when it can expect to
receive the next pulse. This means that we can block the input to
the demodulator circuit until shortly before such a moment. This is
done by the Window Select block. In the experiment this circuit
opens the receiver input at the time $t'_{n+1}=t_n+\beta F(V_n)$ by
Window Control pulses. The input stays open until the decoder is
triggered by the first pulse received. Using such windowing greatly
reduces the chance of the receiver being triggered by noise,
interference or pulses belonging to other users.

\subsection{Parameters mismatch limitations}

It is known that because the synchronization-based chaotic
communication schemes rely on the identity of synchronous chaotic
oscillations, they are susceptible to negative effects of
parameters mismatches. Here we evaluate how precisely the
parameters of our modulator and demodulator have to be tuned in
order to ensure errorless communication over a distortion-free
channel.

Since the information detection in our case is based on the
measurements of time delays, it is important that the modulator and
the demodulator can maintain synchronous time reference points. The
reference point in the modulator is the front edge of the Chaotic
Clock Pulse. The reference point in the demodulator is the front
edge of the Window Control Pulse. Ideally, if the parameters of the
modulator and the demodulator were exactly the same and the systems
were synchronized, then both reference points would be always at
the times $t'_{n+1}=t_n+\beta F(V_n)$, and the received pulse would
be delayed by the time $d$ for $S_{n+1}=0$ and $d+m$ for
$S_{n+1}=1$. In this case, setting the bit separator at the delay
$d+m/2$ would guarantee errorless detection in a noise-free
environment.

\begin{figure}[h]
\begin{center}
\leavevmode
\hbox{%
\epsfxsize=7.2cm
\epsffile{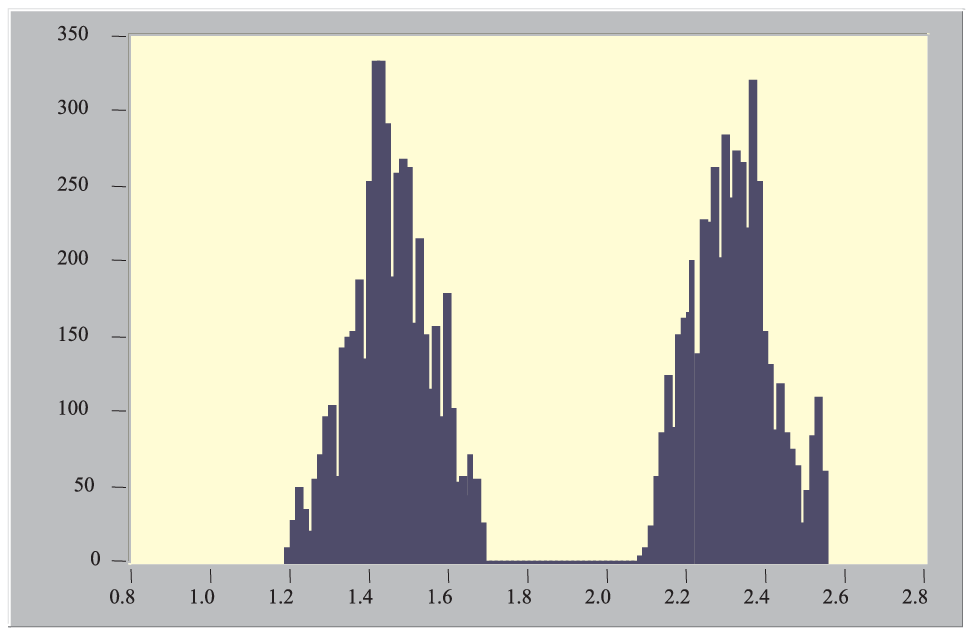}}
\hbox{
\epsfxsize=7.2cm
\epsffile{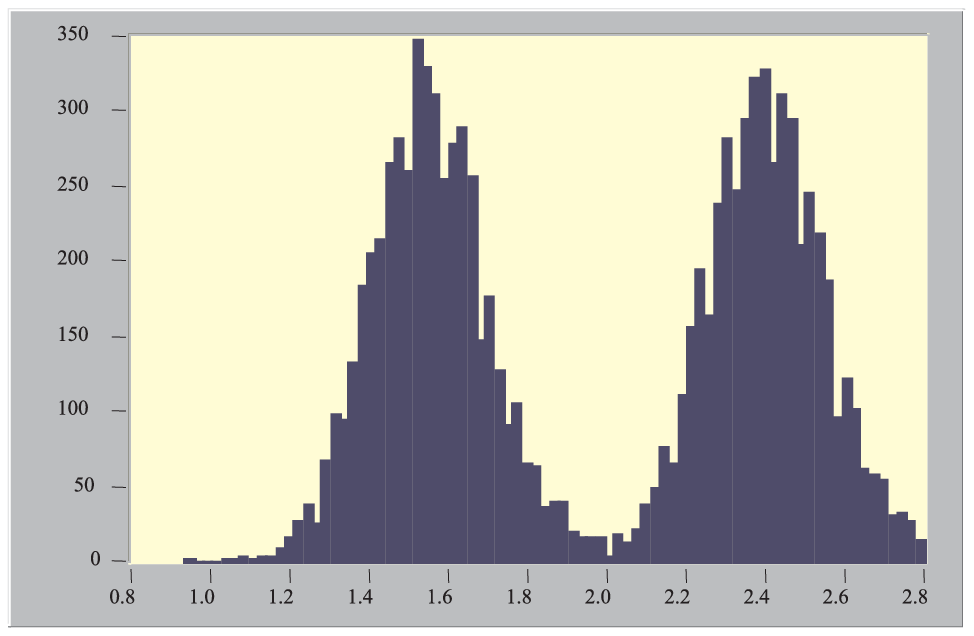}}
\end{center}
\caption{Histograms of the fluctuations of the received pulse
positions with respect to the receiver reference point: noise-free
channel- (top) and channel with WGN $E_b/N_o\sim18$dB
- (bottom).}
\label{hist}
\end{figure}

In an analog implementation of a chaotic pulse position
modulator/demodulator system, the parameters of the circuits are
never exactly the same. Therefore, the time positions $t'^{(M)}_n$
and $t'^{(D)}_n$ of the reference points in the modulator and the
demodulator chaotically fluctuate with respect to each other. Due
to these fluctuations the position of the received pulse,
$t_n=t'^{(M)}_n+ d+S_n$, is shifted from the arrival time predicted
in the demodulator, $t'^{(D)}_n + d+S_n$. The errors are caused by
the following two factors. First, when the amplitude of
fluctuations of the position shift is larger than $m/2$, some
delays for ``0''s and ``1''s overlap and cannot be separated.
Second, when the fluctuations are such that a pulse arrives before
the demodulator opens the receiver input ($t_n<t'^{(D)}_n$), the
demodulator skips the pulse, loses synchronization and cannot
recover the information until it re-synchronizes. In our
experimental setup the parameters $\beta_{M,D}$ were tuned to be as
close as possible, and the nonlinear converters were built using
$1\%$ components. The fluctuations of the positions of the received
pulses with respect to the Window Control pulse were studied
experimentally by measuring time delay histograms.
Figure~\ref{hist} presents typical histograms measured for the case
of noise-free channel and for the channel with noise when
$E_b/N_o\sim18$dB.

Assuming that systems were synchronized up to the $(n-1)$-st pulse
in the train, the fluctuations of the separation between the
reference time positions equals
\begin{eqnarray}\label{delta}
&\Delta_n\equiv t'^{(D)}_n-t'^{(M)}_n= \nonumber\\
&\beta_D
F_D(\beta_D^{-1}T_{n-1})-\beta_M F_M(\beta_M^{-1}T_{n-1}),
\end{eqnarray}
where indices $D$ and $M$ stand for demodulator and modulator
respectively. As it was discussed above, in order to achieve
errorless detection, two conditions should be satisfied for all
time intervals in the chaotic pulse train produced by the
modulator. These conditions are the synchronization condition,
$\{\Delta_n\}_{max}<d$, and the detection condition
$\{|\Delta_n|\}_{max}<m/2$. As an example we consider the simplest
case where all parameters of the systems are the same except for
the mismatch of the parameter $\alpha$ in the nonlinear function
converter, see Eq.(\ref{nlf}). Using Eq.(\ref{nlf}) and
Eq.(\ref{delta}) the expression for the separation time can be
rewritten in the form
\begin{equation}\label{delta_f}
\Delta_n= (\alpha_D - \alpha_M )\beta f(\beta^{-1}T_{n-1}).
\end{equation}
It is easy to show that the largest possible value of the
nonlinearity output $f(~)$, which can appear in the chaotic
iterations of the map, equals to 5V. Note that in the chaotic
regime only positive values of $f(~)$ are realized. Therefore, if
conditions
\begin{equation}\label{delta_f}
\beta(\alpha_D - \alpha_M ) < d/5V \mbox{~ and ~}
 2\beta|\alpha_D -\alpha_M| < m/5V.
\end{equation}
are satisfied and there is no noise in the channel, then
information can be recovered from the chaotic pulse train without
errors.

\begin{figure}[h]
\begin{center}
\leavevmode
\hbox{%
\epsfxsize=7.0cm
\epsffile{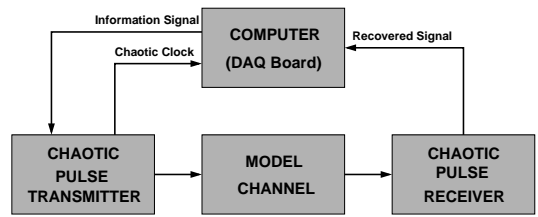}}
\end{center}
\caption{Diagram of the experiment.}
\label{experiment}
\end{figure}

\section{Experimental Setup and Results}

\begin{figure}
\begin{center}
\leavevmode
\hbox{%
\epsfxsize=8.5cm
\epsffile{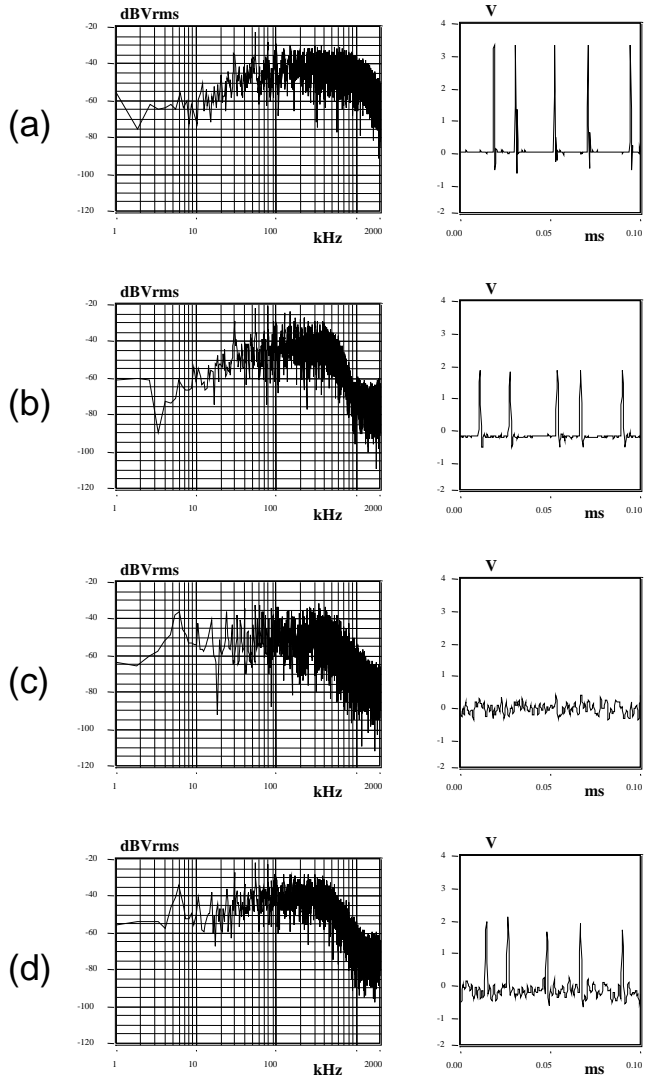}}
\end{center}
\caption{Spectra and waveforms of signals in the channel: $(a)$ --
transmitter output; $(b)$ -- filtered transmitter output; $(c)$ --
filtered noise; $(d)$ -- the received signal.}
\label{signals}
\end{figure}

In our experiment (Fig.\ref{experiment}) we used a computer with a
data acquisition board as the data source, triggered by the chaotic
clock from the transmitter. We also used the computer to record the
pulse displacement from the demodulator subtractor for every
received pulse. This value was used to decode the information for
the bit error rate analysis.  The model channel circuit consisted
of WGN generator and a bandpass filter with the pass band
1kHz-500kHz. The pulse duration was 500ns. The distance between the
pulses varied chaotically between 12$\mu$s and 25$\mu$s. This
chaotic pulse train curried the information flow with the average
bit rate $\sim$60kb/sec. The amplitude of pulse position
modulation, $m$, was 2$\mu$s. The spectra of the transmitter
output, noise and the signal at the receiver are shown in
Fig.\ref{signals}.

\begin{figure}
\begin{center}
\leavevmode
\hbox{%
\epsfxsize=8.0cm
\epsffile{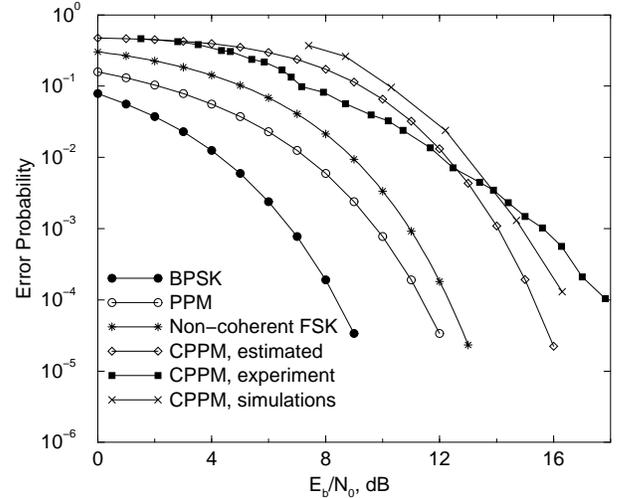}}
\end{center}
\caption{Error probabilities of ideal BPSK, non-coherent FSK, and ideal
PPM systems compared to the performance of the CPPM system.}
\label{results}
\end{figure}

We characterize the performance of our system by studying the
dependence of the bit error rate on the ratio of energy per one
transmitted bit to the spectral density of noise, $E_b/N_0$. This
dependence is shown in Fig.\ref{results}, where it is compared to
the performance of more traditional communication schemes, BPSK,
PPM, and non-coherent FSK.

We were also able to analytically estimate the performance of our
system assuming perfect synchronization. The corresponding curve is
also shown in figure Fig.\ref{results}. At high noise levels the
seemingly better performance of the experimental device compared
with the analytical estimate is in part due to the crudeness of the
analytical model, and in part due to the fact that at high noise
level the noise distribution deviates from Gaussian. In the region
of low noise the deviation of the experimental performance from the
analytical estimate is probably due to the slight parameter
mismatch between the transmitter and the receiver.

Discussing chaos-based communication systems, one may notice a
potential disadvantage common to all such schemes.  Most
traditional schemes are based on periodic signals and systems where
the carrier is generated by a stable system. All such systems are
characterized by zero Kolmogorov-Sinai entropy
$h_{KS}$\cite{Stojanovski97inls}: in these systems without any
input the average rate of non-redundant information generation is
zero. Chaotic systems have positive $h_{KS}$ and continuously
generate information.  Even in the ideal environment, in order to
synchronize two chaotic systems, one must transmit an amount of
information per unit time that is equal to or larger than the
$h_{KS}$\cite{Stojanovski97inls}.  Although our detection method
allows some tolerance in the synchronization precision, the need to
transmit extra information to maintain the synchronization results
in an additional shift of the actual CPPM performance curve
relative to the case when ideal synchronization is assumed.  Since
the numerical and experimental curves in Fig.\ref{results} pass
quite near the analytical estimate that assumes synchronization,
the degradation caused by non-zero Kolmogorov-Sinai entropy does
not seem to be significant.

Although CPPM performs worse than BPSK, non-coherent FSK and ideal
PPM, we should emphasize that ($i$) this wide band system provides
low probability of intercept and low  probability of detection;
($ii$) improves the privacy adding little circuit complexity
($iii$) to our knowledge, this system performs exceptionally well
compared to other chaos-based covert communication
schemes\cite{Chen98inls}; ($iv$) there exist a multiplexing
strategy that can be used with CPPM \cite{Torikai98inls} ($v$)
compared to other impulse systems, CPPM does not rely on a periodic
clock, and thus can eliminate any trace of periodicity from the
spectrum of the transmitted signal. All this makes CPPM attractive
for development of chaos-based cloaked communications.

This research was sponsored in part by the ARO, grant
No.~DAAG55-98-1-0269 and in part by the U.S. Department of Energy,
Office of Basic Energy Sciences, under grant DE-FG03-95ER14516 .

\end{document}